# *Universality of pseudogap and emergent order in lightly doped Mott insulators*

I. Battisti[1]*, K.M. Bastiaans[1]*, V. Fedoseev[1], A. de la Torre[2,3], N. Iliopoulos[1], A. Tamai[2], E.C. Hunter[4], R.S. Perry[5], J. Zaanen[1], F. Baumberger[2,6], M.P. Allan[1¶]

[1] *Leiden Institute of Physics, Leiden University, Niels Bohrweg 2, 2333 CA Leiden, The Netherlands*
[2] *Department of Quantum Matter Physics, University of Geneva, 24 Quai Ernest-Ansermet, 1211 Geneva 4, Switzerland*
[3] *Department of Physics, California Institute of Technology, Pasadena, California 91125, USA*
[4] *School of Physics and Astronomy, The University of Edinburgh, James Clerk Maxwell Building, Mayfield Road, Edinburgh EH9 2TT, United Kingdom*
[5] *London Centre for Nanotechnology and UCL Centre for Materials Discovery, University College London, London WC1E 6BT, United Kingdom*
[6]*Swiss Light Source, Paul Scherrer Institute, CH-5232 Villigen PSI, Switzerland*
* *these authors contributed equally to this work;* ¶ *allan@physics.leidenuniv.nl*

**It is widely believed that high-temperature superconductivity in the cuprates emerges from doped Mott insulators[1]. When extra carriers are inserted into the parent state, the electrons become mobile but the strong correlations from the Mott state are thought to survive; inhomogeneous electronic order, a mysterious pseudogap and, eventually, superconductivity appear. How the insertion of dopant atoms drives this evolution is not known, nor whether these phenomena are mere distractions specific to hole-doped cuprates or represent genuine physics of doped Mott insulators. Here, we visualize the evolution of the electronic states of $(Sr_{1-x}La_x)_2IrO_4$, which is an effective spin-½ Mott insulator like the cuprates, but is chemically radically different[2,3]. Using spectroscopic-imaging STM, we find that for doping concentration of x≈5%, an inhomogeneous, phase separated state emerges, with the nucleation of pseudogap puddles around clusters of dopant atoms. Within these puddles, we observe the same electronic order that is so iconic for the underdoped cuprates[1,4–9]. Further, we illuminate the genesis of this state using unique possibilities on these samples. At low doping, we find evidence for deeply trapped carriers, leading to fully gapped spectra, which abruptly collapse at a threshold of around 4%. Our results clarify the melting of the Mott state, and establish phase separation and electronic order as generic features of doped Mott insulators.**

A core mystery of condensed matter physics is how the rigid arrangement of electrons in Mott insulators loosens when inserting electrons or holes, and how this leads to exotic states inside the Mott gap (Fig. 1). In the cuprate high-temperature superconductors, this process might be the cause of their poorly understood, remarkably complex behavior. Most prominent is the formation of a pseudogap and a variety of inhomogeneous electronic orders[1,4–9], sometimes described as intertwined[10]. This phenomenology has often been assumed (but not verified) to be generic of melting spin-½ Mott physics and not just cuprate-specific. In this Letter we show that an inhomogeneous electronic phase separation as well as a local glassy, stripy charge order exist in a chemically completely different two-dimensional Mott insulator, revealing an universality of these emergent phenomena.

To this end, we create $(Sr_{1-x}La_x)_2IrO_4$ samples with a range of different Lanthanum doping concentrations $x$, $0<x<6\%$ (see SI, Section 1). This material can be seen as a quasi two-dimensional, electron-doped Mott insulator similar to the cuprates despite a very different chemical make-up. The $x=0$ parent material consists of alternating $IrO_2$ and $SrO$ planes, such that oxygen octahedra form around each iridium atom. The five valence electrons in the outer Ir $5d$ shell are split by crystal field and strong spin-orbit coupling to form a filled $J_{eff,3/2}$ and a half filled $J_{eff,1/2}$ band. The moderate on-site repulsion $U$ is then sufficient to open a Mott gap in the $J_{eff,1/2}$ band, making $Sr_2IrO_4$ an effective spin-½ Mott insulator (Fig. 1a,b)[3]. In contrast to hole doping in the cuprates, which is often interstitial oxygen doping, $La^{3+}$ substitutions on the Sr site provide electron-doping for the iridates (Fig. 1c,d). The resulting electronic structure is sometimes predicted to become high-temperature superconducting upon sufficient doping[11,12].

Figures 1e,f depict typical topographs on atomically flat, SrO terminated surfaces for different doping concentrations; the SrO lattice is visible with lattice constant $a_0=3.9$Å and the white squares identify the positions of La dopant atoms in the same layer[13]. The ability to identify the dopant positions easily with atomic precision on the topographs (e.g. in contrast to the cuprates[14,15]) is key to this investigation, as it allows to precisely localize dopant atoms, even when samples show micrometer variations of the dopant concentration (SI, Sections 3,4). For this study, we investigate surfaces with local doping concentrations of 2.1%, 2.2% (Fig. 1e), 2.3%, 3.7%, 4.8%, 5.0% (Fig. 1f), 5.2%, 5.5% (Fig. 3,4) in order to obtain a full overview of the doping evolution (Fig. 5).

We start our discussion with the very lightly doped samples that are deep in the Mott phase. A typical topograph of a sample with 2.2% dopant concentration is shown in Figure 1e. In all our measurements, this doping level yields a clear Mott gap, as shown in Figure 2a. The shape of the gap is reminiscent of STM spectra of cuprate parent materials[5–7,16]. We describe in the SI, Section 5, how the poor screening in lightly doped Mott insulators leads to an additional potential that decays inside the sample, commonly called tip-induced-band-bending[17,18], yielding an apparent energy gap larger than true Mott gap, $\Delta^{app}_{Mott}>>\Delta_{Mott}$, which makes the gap roughly consistent with optical measurements[19].

To investigate how the Mott state reacts when dopant atoms are inserted, we acquire atomic-scale *Mott maps,* i.e. the magnitude of the Mott gap as a function of location, $\Delta_{Mott}(\mathbf{r})$, while measuring the dopant positions on the atomic scale. Each Mott map is extracted from a set of differential conductance spectra measured on a grid $(r_x,r_y)$. Figure 2a shows the result on a 2.2% sample; the dopant atoms are marked by green dots. La dopants do not significantly change the Mott gap size in their close vicinity; instead, they induce or pin long wavelength arrangement of varying Mott gap. We interpret these nanoscale arrangements as the first of a series of orders that appear upon doping. The most surprising observation, however, is the total lack of in-gap states, despite the presence of dopants – a mystery to which we shall return towards the end of this Letter.

The pure Mott state described thus far is not sustained at doping levels above ~5%. At that point, we discover an abrupt transition to a strikingly inhomogeneous electronic structure: a phase separated Mott/pseudogap landscape[20–22]. Some regions still exhibit a pure Mott gap; in contrast to the very low doping samples, now the Fermi level is pinned closer to the bottom of the upper Hubbard band, (blue curve in Fig. 3a), as expected for a Mott insulator doped with free carriers (similar to electron doped bi-layer iridate[23] and opposite to hole doped cuprates[6]). Additionally, there are regions where we measure electronic states inside the Mott gap (red curve in Fig. 3a). Here, the spectra are remarkably similar to the pseudogap in the cuprates[4–6], with a gap value of around 70-300meV, in rough agreement with recent photoemission measurements that extracted the leading edge gap[24–26], and with some spectra showing clear 'coherence peaks' (Fig 3d). We will refer to these regions as 'pseudogap puddles'. They are not randomly distributed, but form around regions with clusters of dopant atoms. Importantly, we do not observe pseudogap puddles in low doped samples, even if a few dopants happen to be close together by chance; a certain threshold in the doping level is needed for the transition to occur.

In order to further analyze this phase separated landscape, it is necessary to establish the spatial distribution of the Mott/pseudogap character. To do so, we introduce a "*Mott parameter*" $M(\mathbf{r})$ by integrating the density-of states inside the putative Mott gap and normalize it by the integrated density-of-states outside the gap, $M(\mathbf{r})=\int_{-350meV}^{+50meV} LDOS(E)dE / \int_{+200meV}^{+500meV} LDOS(E)dE$ (Fig. 3c, inset). This parameter is large when there are states inside the gap, and small when the Mott gap is dominating. Plotting $M(\mathbf{r})$ as a function of the spatial coordinates reveals the nanoscale character of the phase separation, with pseudogap puddles within regions of pure Mott gap. The phase separation is well defined and sharp, in the sense that the transition from pure Mott area to a pseudogap puddle occurs within less than a nanometer (Fig. 3d). This allows us to define a threshold for Mott and pseudogap regions (black contour in Fig. 3c).

We then develop a fitting procedure that is able to fit spectra both in the Mott regions and in the pseudogap puddles. The fitting function includes a phenomenological Mott gap, and additional density of states within that is gapped by a phenomenological pseudogap based on photoemission results[25,26] and commonly used in the cuprates[7,27] (Fig. 3b). This function allows us to simultaneously extract both the pseudopgap $\Delta_{PG}$ (Fig. 3e) and the Mott gap $\Delta_{Mott}$ (SI, Section 7) for ~$10^5$ spectra located in the pseudogap puddles and to calculate the correlations between the two gaps. If the magnetic correlations $J$ in the *t-J* model are directly causing order that manifests the pseudogap, one could expect an *anti*-correlation between $\Delta_{PG}$ and $\Delta_{Mott}$, as $J\sim t^2/U$. Intriguingly, within the puddles, our data show a clear *positive* correlation of 0.31, i.e. the larger the Mott gap, the larger the pseudogap (Fig. 3e, inset). This is evidence that pseudogap and Mott physics are intimately linked[1,28], but suggests that it is not simply the magnetic correlations that cause the pseudogap.

To further test if the cuprate phenomenology is universal to lightly doped Mott insulators, we search for ordered phases on our samples. In the cuprates, it has become very clear that a sizable set of (possibly intertwined) orders coexist, perhaps causing the pseudogap[1]. These include disordered stripy charge arrangements, sometimes referred to as glassy order or charge density waves[1,4–10]. Indeed, we find that the spatial distribution of the pseudogap value, when extracted with atomic precision, reveals a striking tendency for order. The $\Delta_{PG}$ gap maps exhibit glassy, locally unidirectional structures (Fig. 3e and 4a), reminiscent of lightly hole doped cuprates[4–8](see SI, section 6). Glassy charge order is also visible in the density of states right outside the pseudogap, e.g. at -210meV shown in Figure 4b,c. The arrangements consist of bond centered, unidirectional objects of length-scales of 2 to 4 Ir-Ir distances, clearly very disordered on a larger length-scale. These arrangements, like the pseudogap puddles, nucleate around the dopant atom positions (green circles).

Next, we want to elucidate how this inhomogeneous, charge ordered pseudogap state emerges from the fully gapped state at low doping, by using the unique availability of samples with densely spaced doping concentrations in the iridate family. We measure one or more spectroscopic imaging maps with $>10^6$ data-points at each doping concentration, and we analyze each using the methods described above. Figure 5 summarizes the results, illustrating the abrupt nature of the transition. Panel **a** shows the evolution of the averaged spectra in the regions with pseudogap (red) and in those without (blue), panels **b-g** show the phase separation on the respective field of views as defined by the Mott parameter defined above.

At doping concentration below the transition threshold, none of the spectra exhibits any sort of impurity state. Nor is the chemical potential pinned to one of the edges of the Mott gap, as one would expect from a Mott insulator with free carriers from shallow dopant centers. Combined with the fact that the phenomenology of the electronic structure is surprisingly independent of the doping concentration below ~4%, this leads to the question: Where did all the dopant electrons go? We propose the scenario illustrated in Figure 5h. Tightly bound dopant electrons at the dopant locations lead to a putative dopant band (impurity band in semiconductor parlance) inside the gap which keeps the chemical potential around mid-gap. Because the extra electrons from the $La^{3+}$ dopants reside in the upper Hubbard band in the IrO layer, they experience the strong Mott correlations. Consequently, the charging energy $E=e^2/C$ to remove or add electrons (or holes) to the $[La^{3+} + e^-]$ bound state is large enough to split the dopant band and push it outside the Mott gap (Fig. 5h). An equivalent way of describing this, going back to N.F. Mott, is that there is a Mott transition in the dopant band[29]. With increasing doping, screening of the long-range Coulomb interaction by doped carriers leads to an abrupt collapse of the impurity Mott state at doping concentrations lower than one would expect in a static picture. In the cuprates, similar microscopic processes have initially been proposed, but the Mott state is much more fragile: even weak doping of around 2% can destroy the logarithmic divergence in the resistance[30]. This is due to the much smaller energy scales of the trapping in the cuprates; below the transitions, the material behaves similar to a doped semiconductor, with an impurity band close to the energy of the valence band[30]. This is consistent with the later observation that the dopant centers are quite shallow[6]. Based on our results, we predict that LDA+U calculations[3,31] on doped iridates will reveal the trapping of La dopant states to be much deeper than the equivalent states in the cuprates, and that more homogenous samples will reveal a very sharp impurity band metal-insulator transition.

Finally, we would like to go back to the comparison of the electron doped iridate material studied here to the cuprates. Detailed measurements on the cuprates, e.g. $Ca_{2-x}Na_xCuO_2Cl_2$ and $Bi_2Sr_2CaCuO_{8+\delta}$, revealed surprising universalities including the glassy charge order observed in the CuO layer. On first look, $(Sr_{1-x}La_x)_2IrO_4$ seems to be a very different beast: electron instead of hole doping, Ir instead of Cu, $5d^5$ instead of $3d^9$. However, our data clearly shows that the physics of electronic order and the pseudogap are not specific to the cuprates but generic to lightly doped Mott insulators, and we believe that the interplay between dopants and order seen here holds for the cuprates as well. By extension, we can expect $(Sr_{1-x}La_x)_2IrO_4$ to become a high-temperature superconductor with only slightly higher doping concentration.

— — —

*Acknowledgements:* We thank J. Aarts, J.C. Davis, M.H. Hamidian, T. van Klingeren, Jhinhwan Lee, M. Leeuwenhoek, V Madhavan, F.M. Massee, K. van Oosten, J. van Ruitenbeek, S. Tewari, G. Verdoes and J.J.T. Wagenaar for valuable discussions. We acknowledge funding from the *Netherlands Organization for Scientific Research* (NOW/OCW) as part of the *Frontiers of Nanoscience program* and the *Vidi talent scheme,* and from the *Swiss National Science Foundation* (200021-146995).

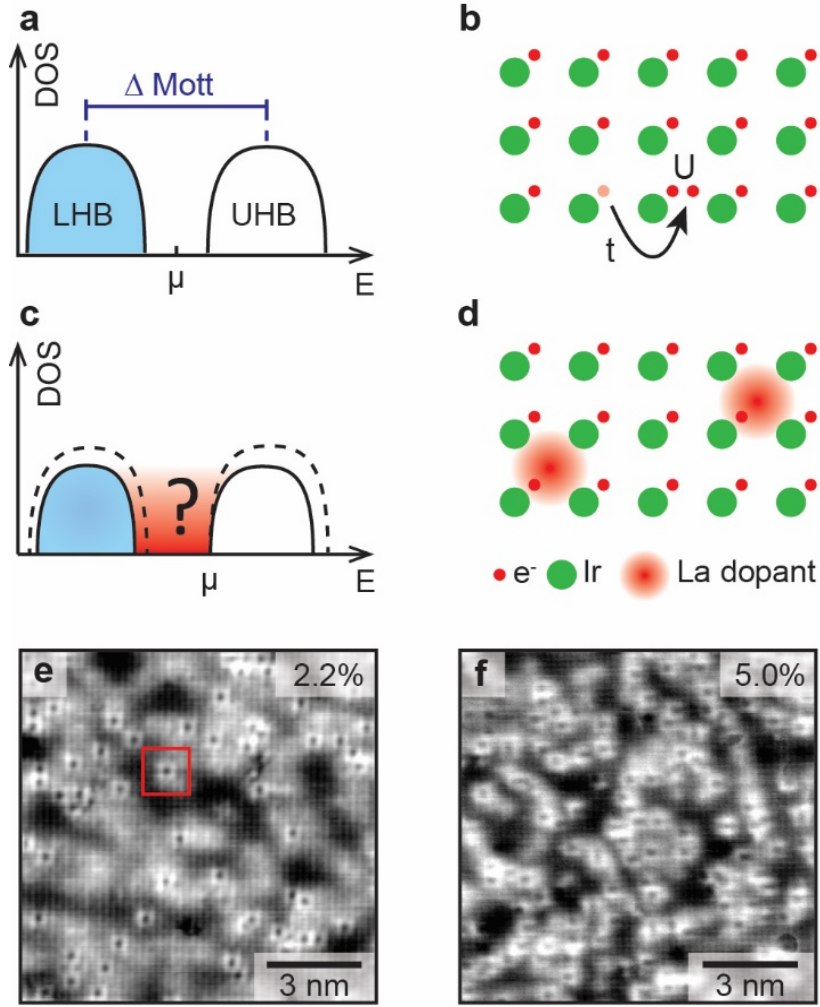

**Figure 1.** The lightly doped effective Mott insulator $(Sr_{1-x}La_x)_2IrO_4$. **a**, In Mott insulators, the electronic states split into a lower and upper Hubbard band separated by the Mott gap. **b**, The gap is caused by the on site interaction $U$ that prevents electrons from hopping from site to site with energy gain $t$. **c,d,** When doping with charge carriers, new states move into the Mott gap in unknown ways. **e**, Atomically resolved topograph with a doping concentration 2.2% (see SI, Section 2,3). The $La^{3+}$ dopant atoms are readily identified as dark spots surrounded by brighter atoms (red square)[13]. **d**, Atomically resolved topograph with a doping concentration of 5.0%. C.f. SI, Section 2 for more topographic images, and Section 3 for information on how we count the dopant atoms with precision of ±0.7%.

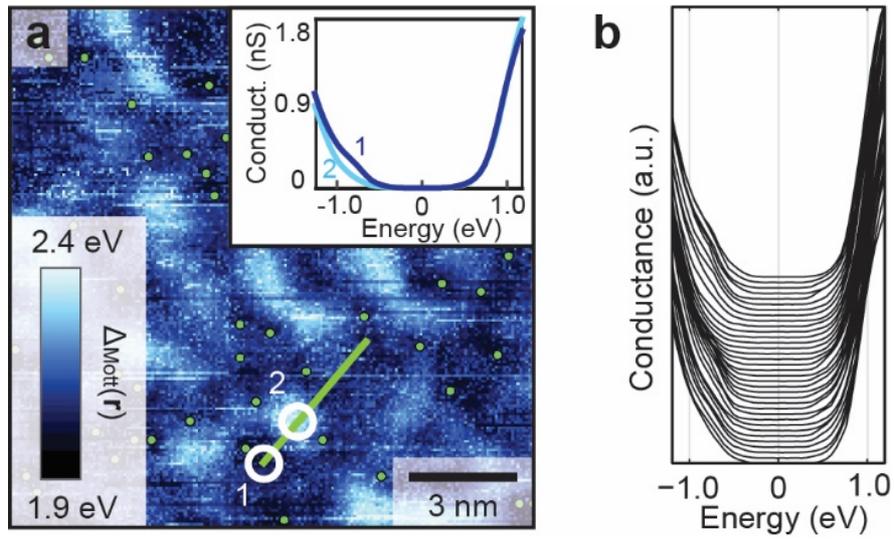

**Figure 2.** The electronic structure of $(Sr_{1-x}La_x)_2IrO_4$ at low doping. **a**, Mott gap map $\Delta_{Mott}$ of a sample with a doping concentration of x=2.2%. Position of dopant atoms are indicated by green circles. The inset shows local density of states spectra averaged inside the white circles. An additional density of states at around -0.8 V is visible in the low gap regions, and might be related to the impending collapse of the impurity Mott state discussed later. **b**, Local density of states spectra along the green line in a, each with an offset on the vertical axis.

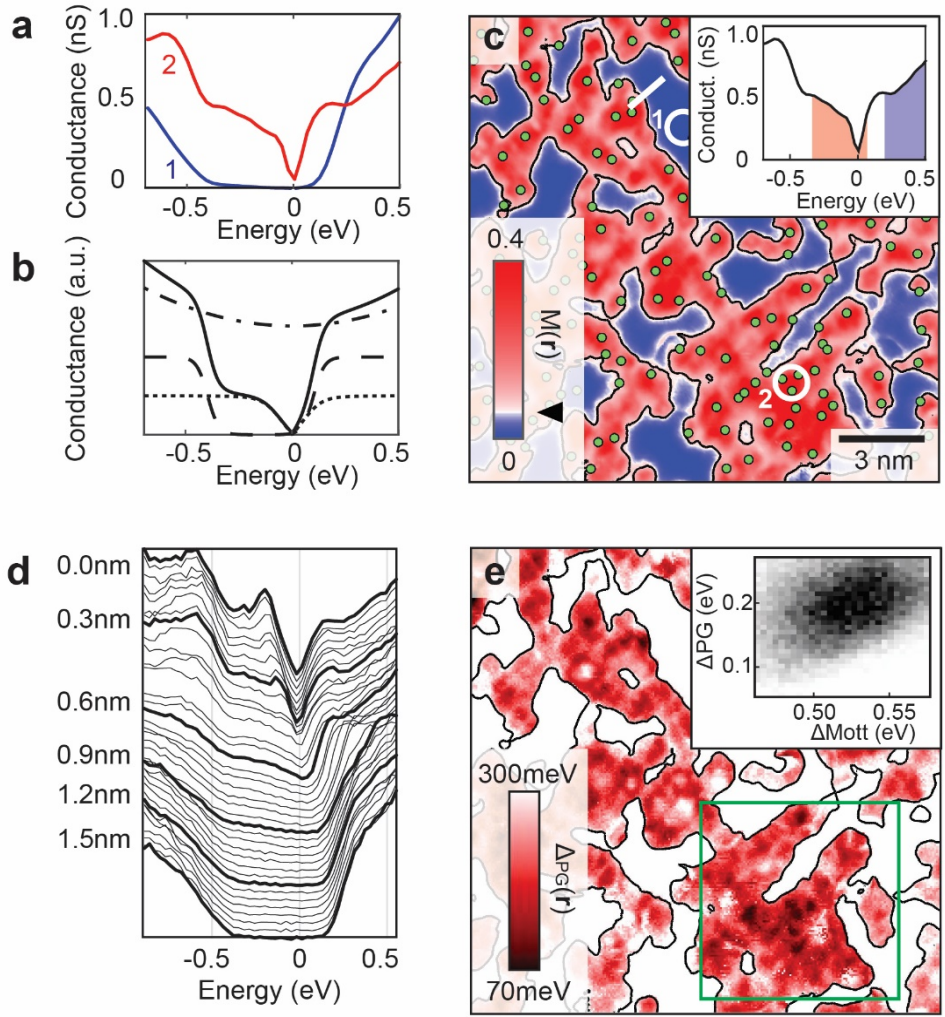

**Figure 3.** Phase separated Mott/pseudogap electronic structure at 5.5% doping. **a,** Different spectra in the phase separated region: Mott like spectrum (blue), with the chemical potential pinned to the UHB, and mixed Mott/pseudogap spectrum (red). The spectra are the average of 180 spectra inside the white circles in panel c. **b,** Phenomenological fit function to simultaneously extract both the Mott and pseudogap size. It consists of a density of states (dot-dashed) multiplied with Mott gap (dashed) plus states inside the Mott gap with a v-shaped pseudogap (dotted). See SI, Section 6 for details. **c,** The Mott parameter as defined in the text identifies pseudogap puddles (red) and pure Mott regions (blue). Green circles indicate La dopant locations. The triangle on the colorbar indicates the value of the black contour. Inset, definition of the Mott parameter: the integrated DOS inside Mott gap (red) normalized by the one outside the gap (blue). **d,** Local density of states spectra along the white line in c (each corresponding to a single measurement). The separation is sharp in the sense that a Mott spectrum becomes a pseudogap spectrum within roughly a nanometer. **e,** $\Delta_{PG}$ map extracted from the fitting procedure (for $\Delta_{Mott}$, see SI Section 7). The square indicates the region displayed in Figure 4. Inset, the correlation between $\Delta_{Mott}$ and $\Delta_{PG}$.

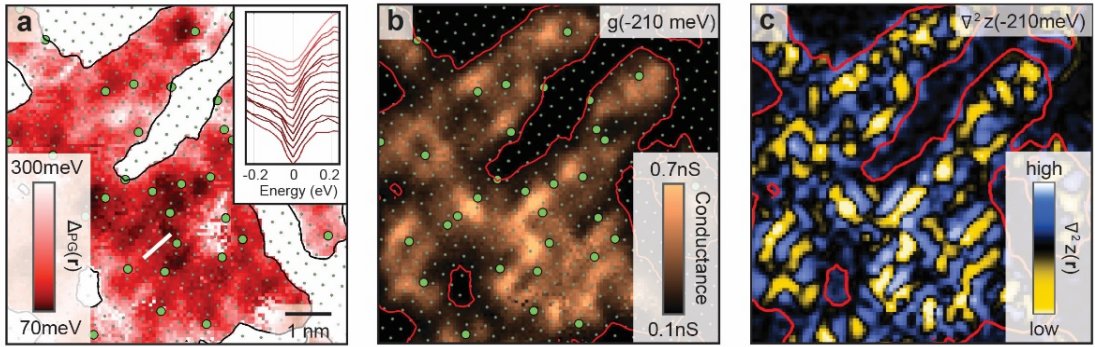

**Figure 4.** Nucleating order. **a**, Map of the pseudogap $\Delta_{PG}$. The small green dots indicate Ir atom location, the larger green circles indicate La dopant locations on the Sr site. Disordered, locally unidirectional patterns are visible. The inset shows a series of spectra along the white line. The color indicates the gap value. **b**, Density of states at -210meV. Glassy order is nucleating around the La dopant atoms. **c**, Laplacian of the ratio map at -210meV.

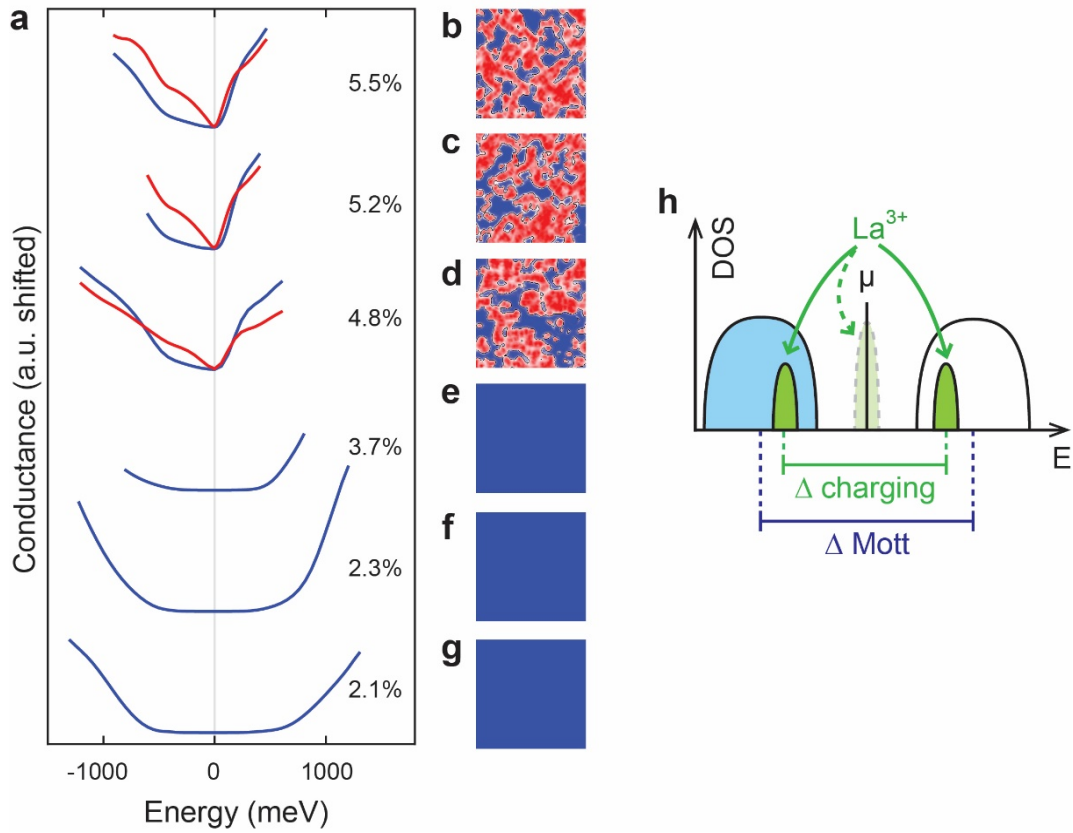

**Figure 5.** The evolution of the electronic structure with increasing dopant atom concentration. **a**, Density of states spectra at different doping levels, each averaged over regions with Mott gap (blue) and pseudogap (red) as defined by the Mott parameter in b-g. At around 5%, the phase separation abruptly starts. **b-g**, Respective maps of the Mott parameter, where blue indicates a pure Mott gap, and red indicates a pseudogap puddle. Increasing doping leads to smaller pure Mott areas. **h,** Schematic image of an impurity Mott transition, with the split impurity band states, (not) including the energy splitting $\varDelta_{charging}$ in (light) green.

# Supplementary Information

"Universality of pseudogap and emergent order
in lightly doped Mott insulators"

I. Battisti[1]*, K.M. Bastiaans[1]*, V. Fedoseev[1], A. de la Torre[2,3], N. Iliopoulos[1], A. Tamai[2], E.C. Hunter[4], R.S. Perry[5], J. Zaanen[1], F. Baumberger[2,6], M.P. Allan[1]¶

[1] Leiden Institute of Physics, Leiden University, Niels Bohrweg 2, 2333 CA Leiden, The Netherlands
[2] Department of Quantum Matter Physics, University of Geneva, 24 Quai Ernest-Ansermet, 1211 Geneva 4, Switzerland
[3] Department of Physics, California Institute of Technology, Pasadena, California 91125, USA
[4] School of Physics and Astronomy, The University of Edinburgh, James Clerk Maxwell Building, Mayfield Road, Edinburgh EH9 2TT, United Kingdom
[5] London Centre for Nanotechnology and UCL Centre for Materials Discovery, University College London, London WC1E 6BT, United Kingdom
[6] Swiss Light Source, Paul Scherrer Institute, CH-5232 Villigen PSI, Switzerland
* these authors contributed equally to this work; ¶ allan@physics.leidenuniv.nl

## 1. Crystal growth and sample characterization

As described in Ref. [1], our $(Sr_{1-x}La_x)_2IrO_4$ single crystals were flux grown from a mixture of off-stoichiometric quantities of $IrO_2$, $La_2O_3$ and $SrCO_3$ in an anhydrous $SrCl_2$ flux. The mixture was heated to 1245°C for 12 hours and cooled at 8°C/hour to 1100°C before quenching to room temperature. The resulting crystals were mechanically separated from the flux by washing with water. The rough La concentration $x$ was determined by energy dispersive x-ray spectroscopy (EDX). Note, however, that this method averages over variations of the doping concentration. We estimate these variations to occur on a length-scale of hundreds of micrometers, and the magnitude to be of the order of a few percentage points. These variations in doping concentrations have to be considered when interpreting resistivity (Figure S1) and magnetization data of the iridate samples.

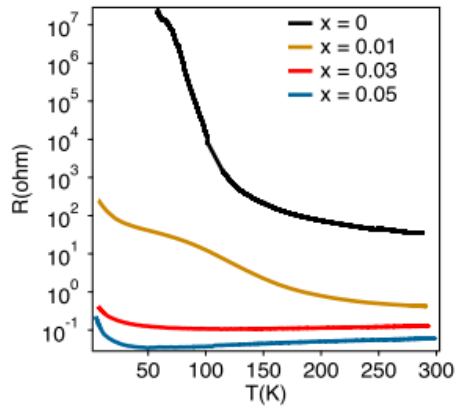

**Fig. S1:** The transition towards metallicity as a function of doping. Figure reproduced from Ref. [1].

## 2. Experimental Methods

The STM experiments are performed with a modified low-temperature, ultrahigh vacuum STM system from Unisoku. The $(Sr_{1-x}La_x)_2IrO_4$ crystals are cleaved *in situ* at temperature T~20 K and base pressure P=2e-10mbar and then transferred immediately into the STM sample stage. All STM results reported in this paper are acquired in cryogenic vacuum at temperatures of either T≈2K or T≈7K (no significant difference has been observed between these temperatures). The STM topographs are taken in the constant current mode, and the dI/dV spectra are collected using a standard lock-in technique with modulation frequency $f$=857 Hz. Importantly, we set each spectrum up at $I_{bias}$ and $V_{bias}$, and then sweep all the voltages. We use mechanically grinded PtIr tips for all the measurements. We always test the spectroscopic and topographic properties of the tips on a crystalline Au(111) surface prepared in situ by Ar ion sputtering and temperature annealing before measuring $(Sr_{1-x}La_x)_2IrO_4$.

The topographs in Fig. 1e,f are set up at -1.2V, 200pA. The map in Fig. 2 is set up at 1.2V, 500pA. The map in Figs. 3,4 is set up at 460mV, 300pA. The maps in Figs. 5 are set up as following: Fig. 5b: 460mV, 300pA; Fig. 5c: -0.4V, 250pA; Fig. 5d: -1.2V, 700pA; Fig. 5e: 0.9V, 220pA, Fig. 5f: 1.2V, 250pA; Fig. 5g: 1.2V, 250pA.

## 3. Set-up effect in topographs

In all samples, we encounter a strong dependence of the topographs appearance on the setup condition, due to the strong electronic inhomogeneity present in $(Sr_{1-x}La_x)_2IrO_4$. Because this inhomogeneity is present up to high bias voltages of order of electron-Volts, it is visible even if the topographs are set up around ~1V. While the topographs do not change strongly upon increasing the absolute value of the bias voltage, we report a strong asymmetry between positive and negative setup. With positive setup (Fig S2a), the dopants have local $C_2$ symmetry, they are more difficult to identify and they sit on a homogeneous background. With negative setup (Fig S2b), they have $C_4$ symmetry and we observe a substantial difference in the contrast between areas with/without dopants.

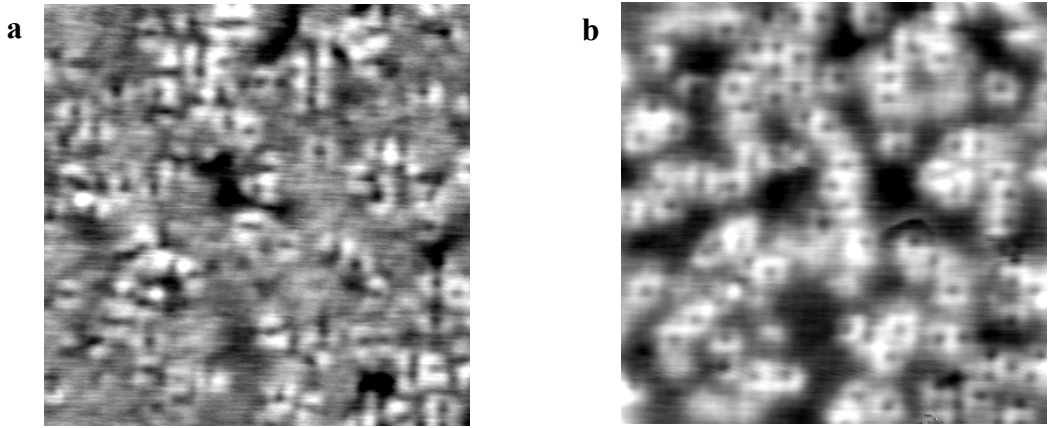

**Fig. S2.** Comparison between two topographs of doping level 5.4% measured on the same area with different setup conditions. (a) Field of view 11.4nm, ($V_{bias}$, $I_{setpoint}$)=(0.5V, 300pA). The dopants have $C_2$ symmetry and the background is homogeneous in contrast. (b) Field of view 11.4nm, ($V_{bias}$, $I_{setpoint}$)=(-0.7V, 800pA). The dopants have C4 symmetry and the background is inhomogeneous.

## 4. Identification of dopant atoms location

Identifying the location of the dopant atoms plays an important role in our analysis, and it is more challenging when the doping level is high. Here, we describe our method to count and localize the dopant atoms.

Given a topograph (Fig. S3a), we superimpose the atomic lattice obtained by Fourier filtering to mark the positions of Sr atoms (green in fig S3b). Using this grid, we identify the La dopant locations (red in figure S3b) which substitute Sr atoms.

In addition, we establish a procedure to find the dopant location in the spectroscopic map measurements. Due to the asymmetry of the setup effect, the topograph contains not only topographic information but also electronic inhomogeneity. While it is possible to identify the dopant positions in the unprocessed topographs (see Fig S3a, b, c), it becomes easier if we cancel part of the electronic inhomogeneity by summing the topograph (Fig S3c) with the integral of a certain number of layers in the opposite bias voltage region (Fig S3d). After filtering the resulting image we obtain what we call the processed image (Fig S3e). There, the impurities are most prominently visible.

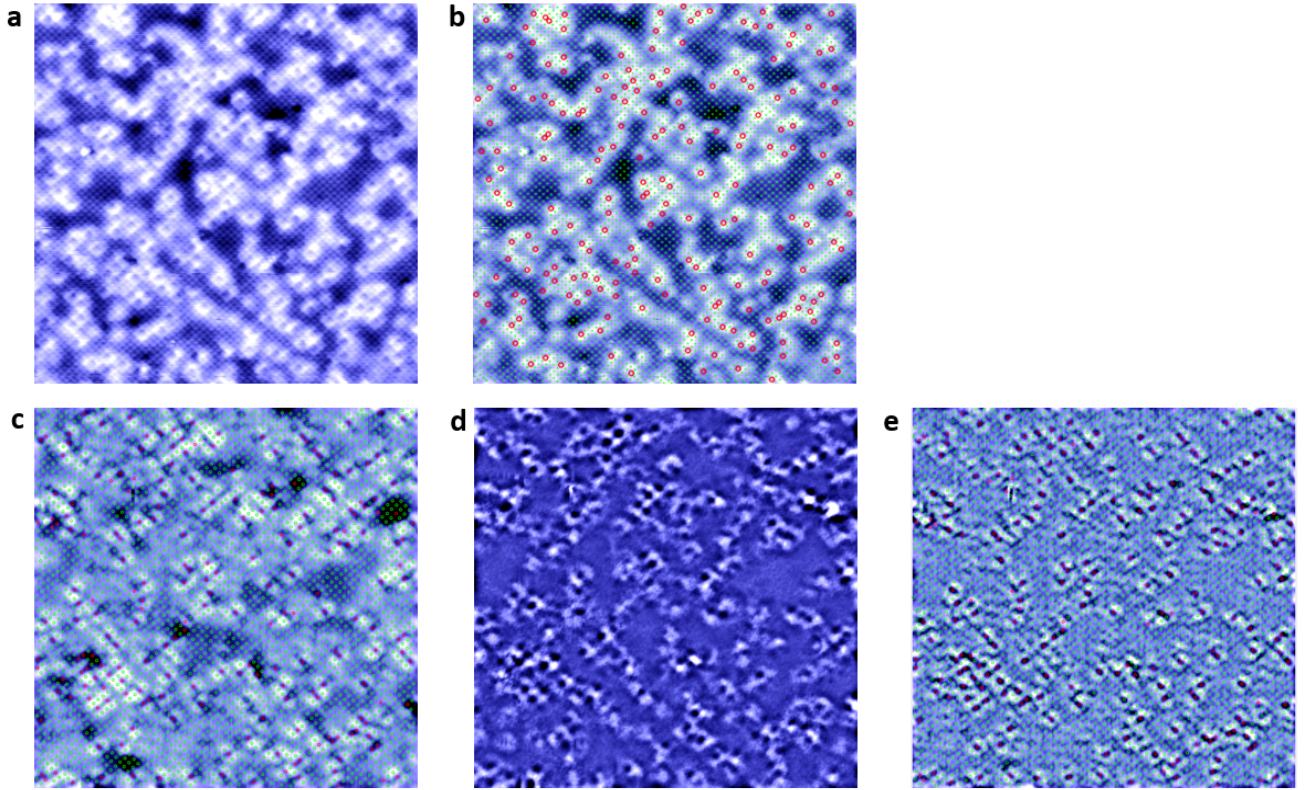

**Fig. S3. a**, Topograph in a field of view of 20 nm, with setup conditions ($V_{bias}$, $I_{setpoint}$)=(-0.75V, 400pA). **b**, Same topograph as in a, with Sr atoms in green and La atoms in red. **c**, Spectroscopic map, field of view 17 nm, ($V_{bias}$, $I_{setpoint}$)=(0.55V, 300pA). Topograph with with Sr atoms in green and La atoms in red. **d**, Integral of conductance layer from -900 meV to -450 meV. **e**, Resulting processed image with Sr atoms in green and La atoms in red.

## 5. Tip-induced band bending

When doing spectroscopy with STM on a metallic sample all the bias voltage drops in the vacuum gap between tip and sample. No voltage drops inside the sample due to the very short screening length of electric field, therefore the bias voltage applied to the sample equals the tip-sample vacuum gap voltage. The applied tip-sample voltage (bias) is equal, on the energy scale, to the distance from the Fermi level of a probed electronic state. On the other hand, in semiconductors [2-6] and Mott insulators [7], the screening length is not negligible. When performing spectroscopy on such samples, the voltage drop inside the sample, so called "tip induced band bending" (TIBB), must be taken into account to retrieve the relative energy of electronic states from the applied bias [2-5].

As our samples are poor conductors with non-metallic resistivity $\rho(T)$ curves (Fig. S1), we can assume that the TIBB should be taken into account. This is confirmed by the observation of semiconductor phenomenology in these samples with SI-STM (gapped density of states, TIBB "bubbles" [3]).

In this section we show that TIBB may result into the apparent band gap being much higher than the real energy gap for our samples.

To get a rough estimate of the voltage drop inside the sample $V_{sample}$ one can consider a homogeneously charged sphere (charges are fixed on the surface) of radius $R$ at a distance $L<<R$ from the sample, seen as a dielectric semi-space with relative permittivity $\varepsilon$ without any free carriers (Fig. S4), and only assuming static fields. Straightforward electrostatic considerations give

$$V_{sample} = V_{bias} \frac{1}{1 + \varepsilon \frac{L}{R}}$$

*Equation 1*

Where $V_{bias}$ is defined as the voltage drop between the point on the sphere closest to the sample and a point infinitely far from it. As the sphere is located very close to the sample ($L<<R$), the tip charge redistribution must be taken into account. Thus a more realistic configuration is the one with a metallic sphere. Using the method of electrostatic images (a charge on the sphere induces an image charge in the sample and a charge in the sample induces an electrostatic image dipole on the sphere), this configuration can be solved numerically. To compare this solution to (Equation 1), it can be expressed as

$$V_{sample} = V_{bias} \frac{1}{1 + c\varepsilon \frac{L}{R}}$$

Where $c$ is a slow varying function on $\varepsilon$ and $R/L$; for example $c=2.88$ for $\varepsilon=50$, $R/L=40$. These values are chosen as we estimate $L\approx 0.5$nm, $R\approx 15$-$50$nm. To our best knowledge, static relative permittivity of undoped $Sr_2IrO_4$ is not available in literature, therefore we estimate it from [8,9] by taking the averaged values, obtaining $\varepsilon_c \approx 30$ for $E||c$-axis and $\varepsilon_{ab} \approx 100$ for $E||ab$ crystal plane.

Therefore, the charge redistribution on the sphere due to the sample proximity decreases the sample voltage drop $V_{sample}$ via an increase of the electric field in the vacuum gap. This is contrary to a naïve expectation of $V_{sample}$ increasing when charge on the sphere is attracted to the sample, causing *L* to become effectively smaller.

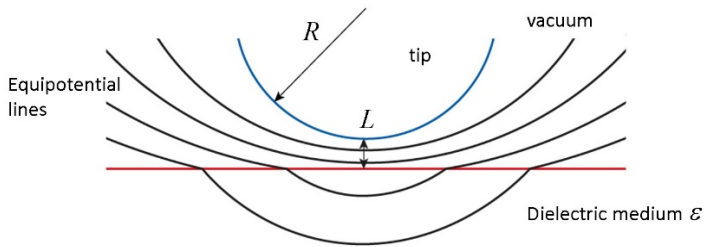

**Fig. S4.** Model to estimate voltage drop inside a non-metallic sample. Tip-sample voltage partly drops in the tip-sample vacuum gap and inside the sample. The tip is modelled by a sphere.

A model closer to an STM setup is a charged hyperbolic metallic surface with radius *R* at the apex and shank opening angle ~30° (2β) located at a distance *L* above the sample (Fig. S5).
This geometry can be solved by finding an appropriate set of a homogeneously charged semi-infinite straight line and a number of point charges between the sample and the line having equivalent boundary conditions to the initial configuration (Fig. S6) [10].

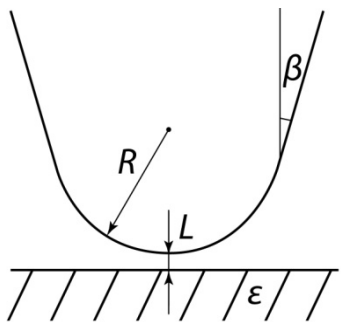

**Fig. S5**. A more appropriate model of the tip with hyperbolic shape.

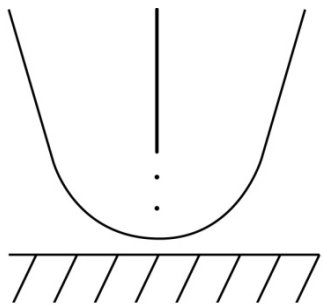

**Fig. S6.** A semi-infinite homogeneously charged line and a few appropriately chosen point charges above a dielectric semi-space have electrostatic fields distribution identical to the hyperbolic tip-sample geometry.

Numerical calculations show that such a geometry results in a ~20% increase of TIBB compared to the spherical metallic tip model. Finally, introduction of free carriers diminishes TIBB as free carriers screen electric field inside the sample.
The uncertainty of tip shape and poorly understood free carrier concentration in doped Mott insulators results in a big error bar for the sample voltage drop. Under assumption of a rather blunt tip and negligible free carrier concentration in lightly doped $Sr_2IrO_4$, TIBB can result in the apparent band gap being few times larger than the real energy gap.

## 6. Disordered stripy pattern

Below, we show a side-by-side comparison of disordered patterns in the conductance layers of iridates with underdoped cuprates. In BSCCO, the patterns are and, based on the limited data that exist, get even more disordered when parts of the sample become more insulating. In our measurements, the pseudogap puddles are smaller than in the measurements on BSCCO and NaCCO, and we thus expect the disorder to be even stronger than in the previously published images.

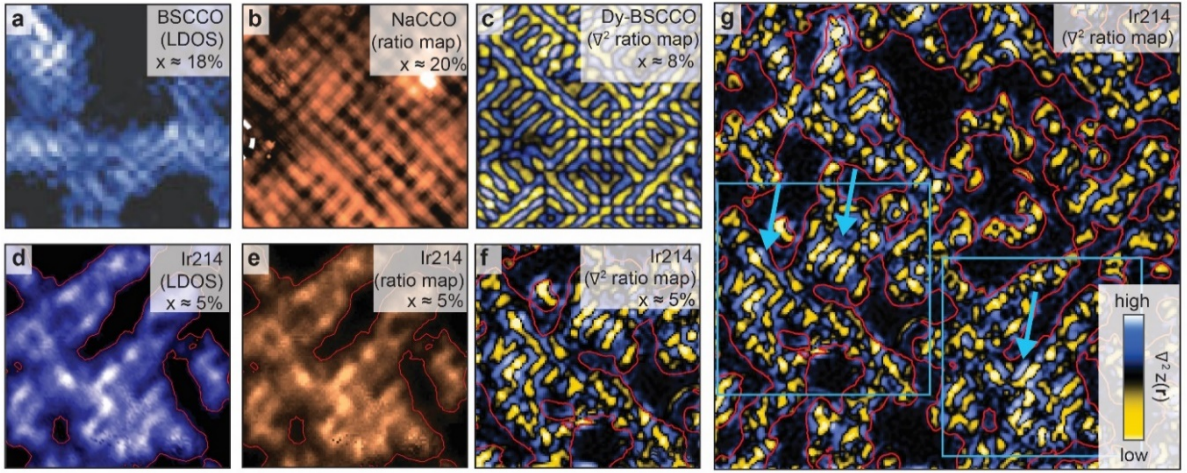

**Fig. S7.** Disordered stripy pattern in the cuprates and the iridates. **a**,**b**,**c**, Disordered stripy pattern in BSCCO and NaCCO seen in the conductance, the ratio map, and the Laplacian of the ratio map, respectively (reproduced from [11, 12, 13]). **d**,**e**,**f**, The corresponding iridate samples exhibit even more disordered patterns likely because they correspond to lower doping. **g,** The larger the pseudogap puddle, the less disordered are the stripy pattern (blue arrows). The blue boxes indicate the regions in d,e,f.

## 7. Fitting procedures

In order to fit both spectra in the Mott region and in the pseudogap puddles, we develop a fitting procedure as follows (Fig. S8). We start with a smooth polynomial background density of states $DOS_{BG}(E) = aE^2 + c$, where $E$ is the energy, and $a$ and $c$ are fitting parameters. Next, we multiply it with a phenomenological Mott gap $\Delta_{Mott}$ consisting of two slightly broadened gap edges, asymmetric around the chemical potential:

$$DOS_{Mott}(E) = \left| \frac{1}{1 + e^{(-E-E_0)/w}} - \frac{1}{1 + e^{(-E+E_0-\Delta_{Mott})/w}} \right|$$

The gap edges are broadened by $w$, $E_0$ is the energy where the upper Hubbard band roughly pins to the chemical potential and $\Delta_{Mott}$ is the size of the Mott gap. We keep the first two parameters fixed ($w$=0.026 eV, $E_0$=0.1 eV), while the size of the Mott gap $\Delta_{Mott}$ is used as a fitting parameter. We then allow for states inside the Mott gap that are gapped by introducing a phenomenological function based on photoemission results and commonly used in the cuprates [14,15]. This part allows for the extraction of the pseudogap value $\Delta_{PG}$.

$$DOS_{PG}(E) = C_0 \left| \frac{E + i\alpha\sqrt{E}}{\sqrt{(E + i\alpha\sqrt{E})^2 - \Delta_{PG}^2}} \right|$$

This function contains two fitting parameters: a scaling factor $C_0$ and the size of the pseudogap $\Delta_{PG}$. We keep α, an effective scattering rate, fixed to 0.2eV$^{0.5}$. The square root in the imaginary part of the self-energy is selected to ensure a rather constant broadening independent of the gap. The resulting model is an excellent fit to all the spectra measured on the highly doped samples, as it can be seen in Fig. S9 where we show 10 randomly chosen spectra with the corresponding fit. Next to the pseudogap energy $\Delta_{PG}$ and the Mott gap energy $\Delta_{Mott}$, our fitting routine also utilizes three other fitting parameters. These are the parameters $a$ and $c$ for the background density of states and the parameter $C_0$ which scales the V-shaped pseudogap function. In Fig. S10 we show the maps of these additional parameters corresponding to the $\Delta_{PG}$ map shown in the main text (Fig. 3e).

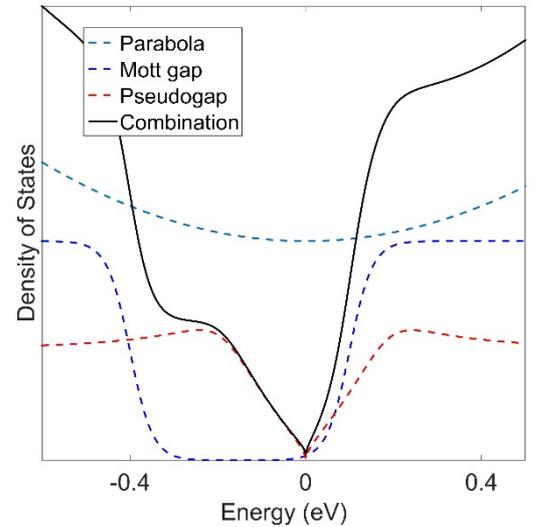

**Fig. S8.** Phenomenological fit function to simultaneously extract both the Mott and pseudogap size. It consists of a polynomial density of states (dashed light blue) multiplied with Mott gap (dashed blue) plus states inside the Mott gap with a v-shaped pseudogap (dashed red).

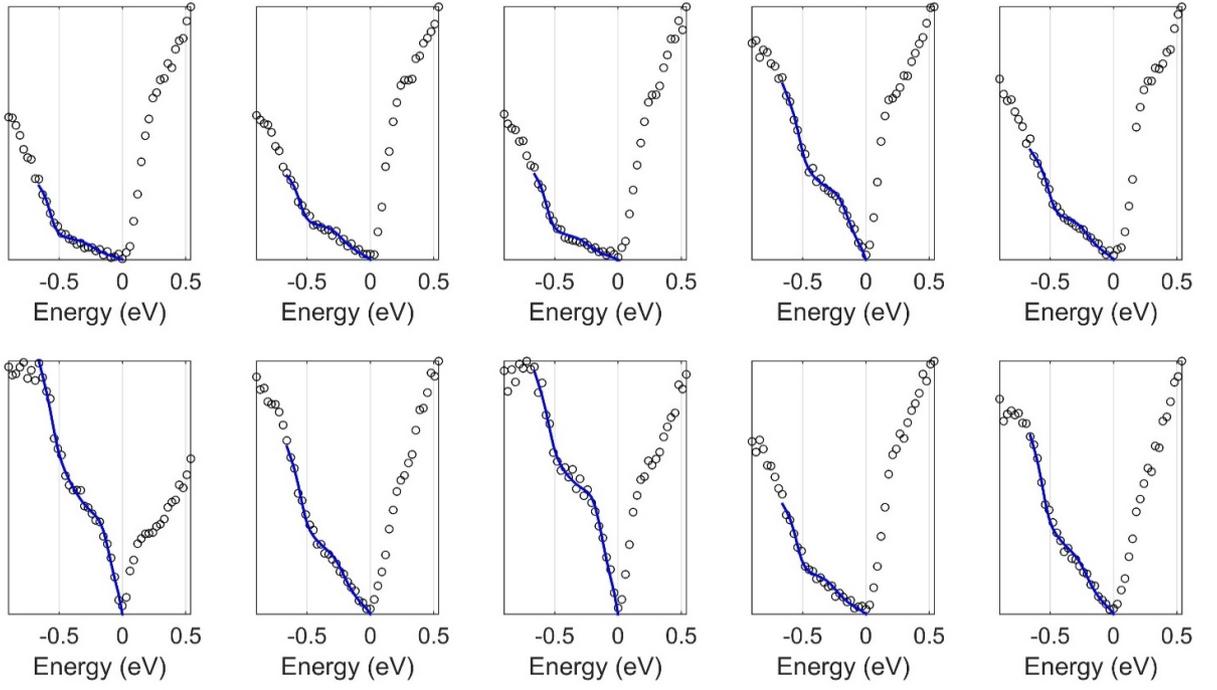

**Fig. S9.** Ten randomly selected curves from the dataset shown in Fig.3 in the text, fitted with the constructed fit function. The open dots represent the data points. The dark blue lines are fits to the data.

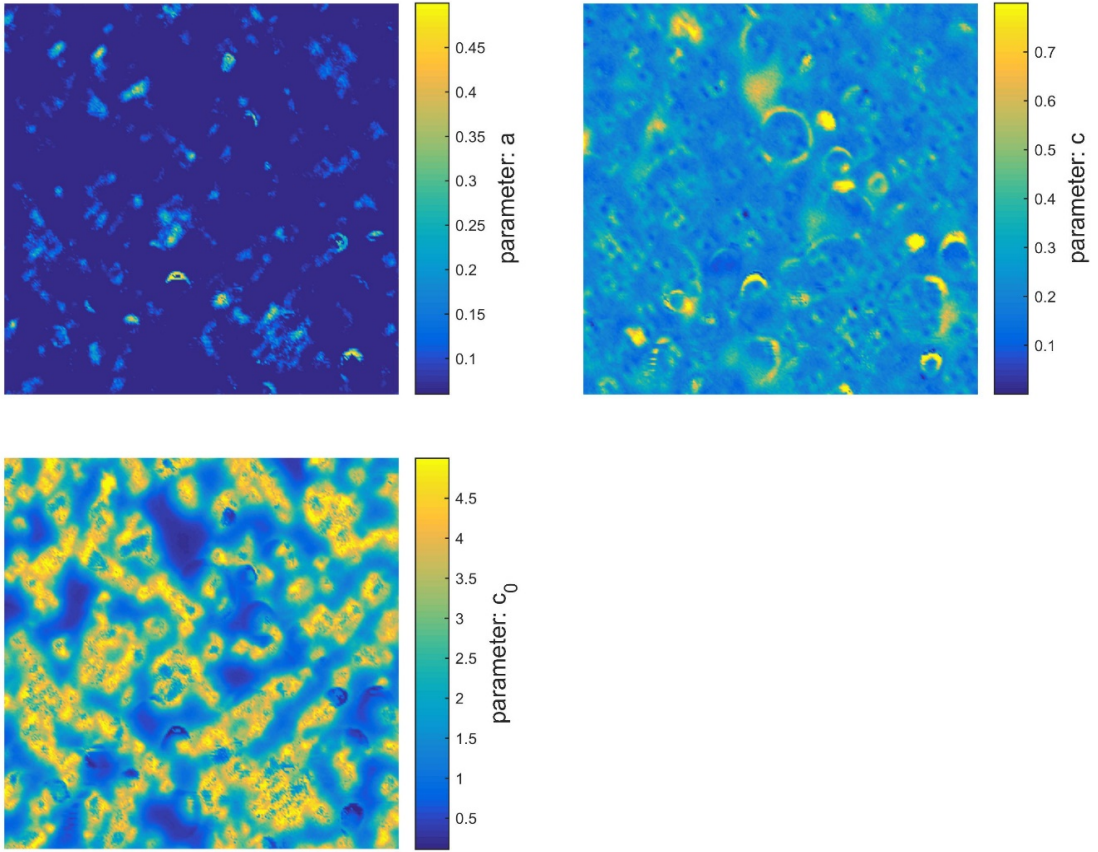

**Fig. S10**. Maps of additional fitting parameters, corresponding to the $\Delta_{PG}$ map shown in Figure 3 of the main text.

## 8. Mott gap map

Using the fitting procedure described in the previous paragraph, we are able to simultaneously extract the value of the Mott gap $\Delta_{Mott}$ and of the pseudogap $\Delta_{PG}$ for each spectrum in the spectroscopic maps. We can then plot Mott gap maps and pseudogap maps as in Fig. S10a,b. The extraction of the values of the two gaps allows us to calculate the correlation between the two gaps within the pseudogap puddles (Fig. S11c).

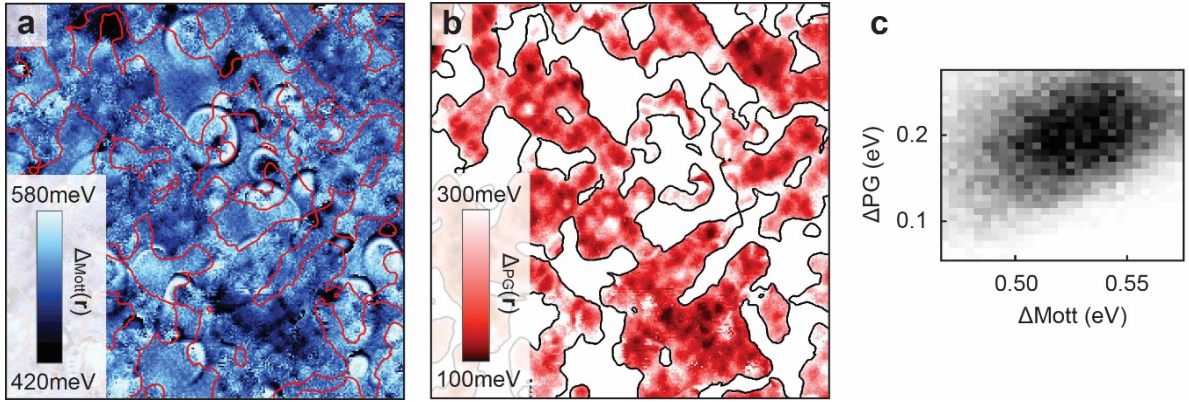

**Fig. S10.** Mott gap map and pseudogap map in a 17 nm field of view region (same measurement as Fig. 3 main text). **a**, Mott gap map. **b**, Pseudogap map. **c**, Correlation between Mott gap and pseudogap.